\input harvmac

\def\epsK{\varepsilon_K}
\def\epe{\varepsilon^\prime/\varepsilon}
\def\gsim{{~\raise.15em\hbox{$>$}\kern-.85em
          \lower.35em\hbox{$\sim$}~}}
\def\lsim{{~\raise.15em\hbox{$<$}\kern-.85em
          \lower.35em\hbox{$\sim$}~}}
\def\Re{{\cal R}e}
\def\Im{{\cal I}m}

\noblackbox
\baselineskip 14pt plus 2pt minus 2pt
\Title{\vbox{\baselineskip12pt
\hbox{hep-ph/0101092}
\hbox{SCIPP-01/03}
\hbox{WIS/01/01-Jan-DPP}
}}
{\vbox{
\centerline{CP Violation and the Scale of}
\smallskip
\centerline{Supersymmetry Breaking}
  }}
\centerline{Michael Dine, Erik Kramer}
\medskip
\centerline{\it Santa Cruz Institute for Particle Physics}
\centerline{\it  University of California, Santa Cruz, CA 95064 }
\centerline{dine@scipp.ucsc.edu, lunenor@physics.ucsc.edu}
\bigskip
\centerline{ Yosef Nir and Yael Shadmi}

\medskip
\centerline{\it Department of Particle
Physics}
\centerline{\it Weizmann Institute of Science, Rehovot 76100, Israel}
\centerline{ftnir@wicc.weizmann.ac.il, yshadmi@wicc.weizmann.ac.il}
\bigskip

\baselineskip 18pt
\noindent
Supersymmetric models with a high supersymmetry breaking scale give,
in general, large contributions to $\varepsilon_K$ and/or to various
electric dipole moments, even when contributions to CP conserving,
flavor changing processes are sufficiently suppressed. Some examples
are models of dilaton dominance, alignment, non-Abelian flavor symmetries,
heavy first two generation sfermions, anomaly mediation and gaugino
mediation. There is then  strong motivation for `approximate CP',
that is a situation where all CP violating phases are small.
In contrast, in supersymmetric models with a low breaking scale it is
quite plausible that the CKM matrix is the only source of flavor and CP
violation. Gauge mediation provides a concrete example.
Approximate CP is then unacceptable. Upcoming measurements of the CP asymmetry
in $B\rightarrow\psi K_S$ might exclude or support the idea of approximate
CP and consequently probe the scale of supersymmetry breaking.

\Date{1/01}

\newsec{Introduction}
CP violation has, until recently, been one of the least explored areas of the
Standard Model. There is little evidence as to whether the smallness of
the CP violation observed in the kaon system is due to flavor suppression,
as is the case for the Kobayashi-Maskawa (KM) mechanism of the Standard Model,
or whether it is due to a suppression of all CP violating phases.
The latter option is known as ``approximate CP''.

Approximate CP could naturally arise if CP is an exact symmetry
of the microscopic theory that is spontaneously broken
\ref\DLM{M.~Dine, R.~G.~Leigh and D.~A.~MacIntire,
Phys.\ Rev.\ Lett.\  {\bf 69}, 2030 (1992) [hep-th/9205011].}%
\ref\CKN{K.~Choi, D.~B.~Kaplan and A.~E.~Nelson,
Nucl.\ Phys.\  {\bf B391}, 515 (1993) [hep-ph/9205202].}.
Indeed, in various string theory compactifications, CP is an exact gauge
symmetry. It is then possible, as we will demonstrate through a simple toy
model, to obtain small CP violating phases if some superpotential couplings
are small; this is entirely natural. In other words, given that in our
experience, most Yukawa couplings are small, approximate CP is a likely
(though certainly not inevitable) consequence of spontaneous
CP violation (SCPV).

The advent of $B$-factories could dramatically change the current picture
of CP violation. The Standard Model predicts large CP-violating asymmetries in
neutral $B$ decays. Measuring small asymmetries would therefore signal new
physics, possibly obeying approximate CP.
Conversely, measuring large asymmetries would rule out approximate CP.
In this paper we therefore examine various supersymmetric extensions
of the Standard Model, in which new sources of CP violation typically
abound, from the point of view of their compatibility with approximate
CP, and their implications for CP violation in the $B$ system.
We will argue that in theories with high scale supersymmetry breaking,
there is typically strong motivation for approximate CP.
In contrast, in theories with low scale supersymmetry breaking,
approximate CP is typically unacceptable.

If the CP asymmetry in $B\rightarrow\psi K_S$
is measured to be very small, supersymmetry with high breaking scale
will be a very likely explanation, the Standard Model will be excluded,
and a supersymmetric extension with low scale breaking will be unlikely.

Supersymmetric extensions of the Standard Model currently face
two classes of constraints from CP violation. One has to do with
CP violation in the $K$ system, or the smallness of $\epsK$.
The second constraint is the smallness of electric dipole moments.
The question of supersymmetric contributions to $\epsK$ is related
to the question of supersymmetric contributions to $\Delta m_K$.
The requirement that these be adequately suppressed already implies that
the squark mass matrices must have some special features.
The supersymmetric contributions to $\epsK$ involve the phases
appearing in these mass matrices, and the mechanism which suppresses
$\Delta m_K$ will have implications for $\epsK$. Supersymmetric contributions
to EDMs, on the other hand, involve the $A$ terms for the sfermions and the
$B$ term for the Higgs, and therefore
depend on the relative phases between these terms,
the gaugino mass, the Yukawa couplings and the $\mu$ term.
Approximate CP is one possible explanation for the smallness
of dipole moments.  Alternatively, they might be small because
of the existence of relations among these terms.
Relations of this type, as we will review, sometimes
arise in models of high scale supersymmetry breaking.
As is well known, even in theories where the $A$ term and
gaugino phases are correlated, the relation between the $\mu$
and $B$ term phases tends to be highly model-dependent.
Alternatively, $A$ term contributions
may be suppressed if the size of the $A$ terms is small,
as typically occurs in models of low energy supersymmetry breaking
(gauge mediation), and might occur in other theories.

In all of our discussion, we will suppose that we are dealing
with a theory which provides an explanation of the suppression
of flavor-changing processes (particularly $\Delta m_K$). There have been
a number of proposals, and we will review them here.
As we will see, theories with high-scale supersymmetry breaking
in which the supersymmetry breaking terms of the Minimal Supersymmetric
Standard Model (MSSM) are generated near the Planck scale, typically predict
$\epsK$ and EDMs values that are too large, if CP-violating phases
are of order one. First, the suppression of flavor changing
couplings (``flavor suppression'') in these theories
is barely sufficient to satisfy the $\Delta m_K$ constraint.
As a result, such theories generically fail to satisfy the $\epsK$
constraint, and approximate CP seems to be a likely ingredient of such theories.
Second, some or all of the phases in the $A$ and/or $B$ terms in
these theories are generally not related to the phases of the Yukawa couplings
and the $\mu$ term. There are then new contributions to EDMs that are too
large, unless these phases are small.
Such theories therefore require approximate CP, and
would then lead to small CP asymmetries in $B$ decays.

On the other hand, in theories of low scale supersymmetry breaking,
such as Gauge Mediation, it is very difficult to accommodate approximate CP.
New contributions to $\epsK$ are flavor suppressed
because the supersymmetry breaking terms are flavor blind.
Therefore, the Standard Model itself has to account for $\epsK$,
and this requires a large KM phase.
Measuring small asymmetries in $B$ decays would therefore
rule out simple models of low energy supersymmetry breaking.

Models in the literature with approximate CP typically predict too small
a value of $\epe$. There is, however, no fundamental reason for this situation
and, as we will show, it is possible to construct models in which $\epe$ is of
the correct size. These models are not generic, and may require some fine
tuning. If the idea of approximate CP finds support in measurements of CP
asymmetries in $B$ decays, the construction of attractive models within
this framework with  $\epe={\cal O}(10^{-3})$ will become an important
challenge.

This paper is organized as follows. In Section~2, we  illustrate
the idea of approximate CP through some simple toy models.
In Section~3, we summarize the CP violation constraints
on supersymmetric models, and the main possibilities for suppressing
flavor violations and hence $\epsK$, emphasizing universality.
We then review the predictions for CP violations in various
models of supersymmetry breaking, starting with low scale models
in Sec~4, and going on to high scale models in Section~5.

\newsec{Approximate CP}
That spontaneous CP violation can be small is perhaps obvious, but it is worth
illustrating the issues with some simple models. We can imagine that CP is
violated by the dynamics which breaks supersymmetry, or by other dynamics,
perhaps associated with flavor physics. The former assumption is the simplest,
and we focus on it here.

We consider an O'Raifeartaigh model with three gauge-singlet fields, $X$, $Y$
and $A$. (We later comment on the coupling to supergravity.) We impose the
following three symmetries:
\eqn\symm{\eqalign{
{\rm CP-symmetry}&:\ \ \ \Phi\rightarrow\Phi^\dagger\ \ (\Phi=X,Y,A);\cr
{\rm R-symmetry}&:\ \ \ R(X)=R(Y)=2,\ \ R(A)=0;\cr
{\rm D-symmetry}&:\ \ \ A\rightarrow-A,\ \ X\rightarrow X,\ Y\rightarrow Y.\cr}}
Then the most general superpotential that is consistent with these symmetries
is given by
\eqn\WwithD{W=X(\lambda_X A^2+\mu_X^2)+Y(\lambda_Y A^2+\mu_Y^2),}
where all couplings are real due to the CP symmetry. For $X$ and $Y$, the
minimum condition reads
\eqn\minXY{\lambda_X\vev{X}+\lambda_Y \vev{Y}=0.}
Thus these two VEVs are not determined classically, but it is natural
to have at the minimum $\vev{X}=\vev{Y}=0$ since this solution preserves
the R-symmetry. To find $\vev{A}$, we define $\lambda^2\equiv\lambda_X^2
+\lambda_Y^2$ and $\mu_\lambda^2\equiv\lambda_X\mu_X^2+\lambda_Y\mu_Y^2$.
For $\mu_\lambda^2>0$, we get
\eqn\minA{A_R\equiv\vev{\Re A}=0,\ \ \
A_I\equiv\vev{\Im A}=\sqrt{\mu_\lambda^2\over\lambda^2}.}
Note that these VEVs preserve a subgroup of the CP and D symmetries
that can be called a CP$^\prime$ symmetry:
\eqn\CPpr{X\rightarrow X^\dagger,\ \ \ Y\rightarrow Y^\dagger,\ \ \
A\rightarrow-A^\dagger.}
Therefore, this model does not have SCPV. Supersymmetry is broken because
$F_X\neq0$ and $F_Y\neq0$. Both $F_X$ and $F_Y$ are, however, real,
consistent with the CP$^\prime$ symmetry.

We modify the model by allowing a small breaking of the D symmetry:
\eqn\breakD{W_{\not D}=A(m_X X+m_Y Y).}
The breaking is small in the sense that the scale of $m_i^2$ is much
smaller than the scale of $\mu_i^2$. It is easy to see that $\vev{X}=\vev{Y}=0$
is still an R-symmetry conserving minimum. As concerns $\vev{A}$, to leading
order in $m^2/\mu^2$ we get
\eqn\minAIR{\eqalign{
A_I\ \approx&\ \sqrt{\mu_\lambda^2\over\lambda^2},\cr
A_R\ \approx&\ -{1\over4}\left[\left({\lambda_X\over\lambda^2}+
{\mu_X^2\over\mu_\lambda^2}\right)m_X+\left({\lambda_Y\over\lambda^2}+
{\mu_Y^2\over\mu_\lambda^2}\right)m_Y\right].\cr}}
Consequently, CP is spontaneously broken. In particular,
\eqn\FwoD{\eqalign{\vev{F_X}\ =&\ \mu_X^2-\lambda_X A_I^2
+iA_I\left(m_X+2\lambda_X A_R\right),\cr
\vev{F_Y}\ =&\ \mu_Y^2-\lambda_Y A_I^2
+iA_I\left(m_Y+2\lambda_Y A_R\right),\cr}}
(and $\vev{F_A}=0$). We learn that $F_X$ and $F_Y$ carry (different) small
phases, of order $m/\mu$.

As a consequence of the complex nature of $F_X F_Y^*$,
supersymmetry breaking terms will carry small phases. For example, there
will be CP violating contributions to squark masses-squared, to $A$-terms
and to gaugino masses:
\eqn\susyb{\eqalign{{\rm squark\ masses:}&\ \ \
\int d^4\theta{\gamma^{ij}\over M^2}X^\dagger Y Q_i^\dagger Q_j+{\rm h.c.};\cr
{\rm A\ terms:}&\ \ \ \int d^2\theta (a_XX+a_YY) H_uQ_i\overline{u}_j;\cr
{\rm gaugino\ masses:}&\ \ \ \int d^2\theta (b_XX+b_YY)W_\alpha^2.\cr}}
So this simple model leads to CP violation, with phases
that can be small as certain parameters in the superpotential, related to the
breaking of a discrete symmetry, become small.

Obviously, we can retain the approximate CP but achieve richer phenomenology
by adding more fields and couplings. An appealing choice is to embed our simple
model in a grand unified framework, with $A$ now an adjoint field. The
$A \rightarrow -A$ symmetry must now be broken by non-renormalizable couplings.
The ratio of the unification scale to the Planck (fundamental) scale could then
be the source of the small CP-violating parameter.

If we couple these models to supergravity, we must face issues of naturalness
connected with the cosmological constant problem.  In particular, the constant
in the superpotential required to cancel the cosmological constant breaks any
would-be R symmetry, and it is difficult to write down models which are the
most general consistent with symmetries which break supersymmetry. This is a
problem in the original Polonyi model. As in that case, one can suppose, for
example, that certain terms in the superpotential are small, and
find a stable or metastable local minimum of the potential.

The model that we presented in this section demonstrates that spontaneous
CP violation can naturally induce small phases in the supersymmetry
breaking F-terms. There is another, very different, mechanism that leads to
approximate CP. It could be that the spontaneous CP breaking leads to VEVs
with phases of order one, but the information about CP violation is
communicated to the observable sector in a way that suppresses the phases in
the low energy effective theory, that is the supersymmetric standard model. The
suppression could be a result of mediation through non-renormalizable terms, as
in the models of ref.
\ref\EyNi{G.~Eyal and Y.~Nir,
Nucl.\ Phys.\ {\bf B528}, 21 (1998) [hep-ph/9801411].},
or of gauge mediation, as in the models of ref.
\ref\BaBa{K.S.~Babu and S.M.~Barr,
Phys.\ Rev.\ Lett.\ {\bf 72}, 2831 (1994) [hep-ph/9309249].}.

\newsec{Constraints on Supersymmetry from CP Violation}
In this section we discuss the constraints on CP violation
in supersymmetric extensions of the Standard Model.
We first briefly review the generic contributions
to CP violating quantities in such models. For arbitrary
supersymmetry breaking terms, these contributions are far too large.
We then go on to enumerate different patterns of scalar
masses that lead to small flavor violations, thus suppressing $\epsK$.
We focus on the case of universal masses, which is a good starting point
for understanding the basic issues associated with approximate CP.

\subsec{Generic CP violating contributions}
The MSSM superpotential is given by
\eqn\suppot{W=Y_{ij}^u H_u Q_{Li} U_{Rj}+
Y_{ij}^d H_d Q_{Li} D_{Rj}+
Y_{ij}^\ell H_d L_{Li}\ell_{Rj}+\mu H_uH_d.}
Supersymmetry breaking in the MSSM appears through the Lagrangian
\eqn\susyb{\eqalign{{\cal L}_{\rm soft}\ =\ -&\ \left(
A_{ij}^u H_u \tilde Q_{Li}\tilde U_{Rj}+
A_{ij}^d H_d \tilde Q_{Li}\tilde D_{Rj}+
A_{ij}^\ell H_d \tilde L_{Li}\tilde \ell_{Rj}+{\rm h.c.}\right)\cr
-&\ BH_uH_d-\sum_{\rm all\ scalars}(m^2_{S})_{ij} A_i\bar A_j-{1\over2}
\sum_{(a)=1}^3\left(m_{1/2}^{(a)}(\lambda\lambda)_{(a)}+{\rm h.c.}\right),}}
where $S=Q_L,D_R,U_R,L_L,\ell_R$, $A$ denotes scalar fields and $\lambda$
gaugino fields.
A typical supersymmetric contribution to $\epsK$ comes from box diagrams
with intermediate gluinos and squarks, giving
\ref\GGMS{F.~Gabbiani, E.~Gabrielli, A.~Masiero and L.~Silvestrini,
Nucl.\ Phys.\  {\bf B477}, 321 (1996) [hep-ph/9604387].}
\eqn\sepsk{\varepsilon_K={5\alpha_3^2\over162\sqrt{2}}{f_K^2m_K\over\tilde m^2
\Delta m_K}\left[\left({m_K\over m_s+m_d}\right)+{3\over25}\right]\Im
\left[(\delta_{12}^d)_{LL}(\delta_{12}^d)_{RR}\right].}
Here
\eqn\defdel{\eqalign{
(\delta_{12}^d)_{LL}=&\
\left({m^2_{\tilde Q_2}-m^2_{\tilde Q_1}\over m^2_{\tilde Q}}\right)
\left|K_{12}^{dL}\right|,\cr
(\delta_{12}^d)_{RR}=&\
\left({m^2_{\tilde D_2}-m^2_{\tilde D_1}\over m^2_{\tilde D}}\right)
\left|K_{12}^{dR}\right|,\cr}}
where $\tilde m$ is the typical scale of the soft supersymmetry breaking
terms, $m_{\tilde Q}^2$ ($m_{\tilde D}^2$) are the masses-squared of the
squark doublets (down-type singlets), and $K_{ij}^{dL}$ ($K_{ij}^{dR}$) are
the mixing angles in the gluino couplings to left-handed (right-handed) down
quarks and their scalar partners. The $\epsK$ constraint then reads
\eqn\epscon{\left({300\ {\rm GeV}\over\tilde m}\right)^2\Im
\left[(\delta_{12}^d)_{LL}(\delta_{12}^d)_{RR}\right]\lsim5\times10^{-8}.}
The $\Delta m_K$ constraint on $\Re\left[(\delta_{12}^d)_{LL}
(\delta_{12}^d)_{RR}\right]$ is about two orders of magnitude weaker.
One can distinguish then three interesting regions for
$\vev{\delta_{12}^d}=\sqrt{(\delta_{12}^d)_{LL}(\delta_{12}^d)_{RR}}$:
\eqn\ranmot{\vev{\delta_{12}^d}\cases{
\gg0.003&excluded,\cr
\gg0.0002\ {\rm and}\ \lsim0.003&viable with small phases,\cr
\ll0.0002&viable with ${\cal O}(1)$ phases.\cr}}
The first bound comes from the $\Delta m_K$ constraint (assuming that the
relevant phase is not particularly close to $\pi/2$). The bounds here apply to
squark masses of order 500~GeV and scale like $\tilde m$. There is also
dependence on $m_{\tilde g}/\tilde m$, which is here taken to be one.

The supersymmetric contribution to the electric dipole moment of the down
quark (which is one of the main sources of $d_N$) can be estimated,
for models in which the $A$ terms are proportional to the SM Yukawa couplings,
$A_{ij}= A Y_{ij}$, to be
\ref\BuWy{W.~Buchmuller and D.~Wyler, Phys.\ Lett.\
{\bf B121}, 321 (1983).}%
\ref\PoWi{J.~Polchinski and M.~B.~Wise,
Phys.\ Lett.\  {\bf B125}, 393 (1983).}%
\ref\FPT{W.~Fischler, S.~Paban and S.~Thomas,
Phys.\ Lett.\  {\bf B289}, 373 (1992) [hep-ph/9205233].}
\eqn\dnsus{d_d=m_d{e\alpha_3 |m_{\tilde g}|\over18\pi\tilde m^4}\left(
|A|\sin\phi_A+\tan\beta|\mu|\sin\phi_B\right),}
where
\eqn\motdn{\phi_A=\arg(A^* m_{\tilde g}),\ \ \
\phi_B=\arg(m_{\tilde g}\mu B^*).}
The $d_N$ constraint reads then
\eqn\dncon{\left({300\ {\rm GeV}\over\tilde m}\right)^2\sin\phi_{A,B}\lsim0.03.}

\subsec{Suppressing FCNCs and $\epsK$}
As in the Standard Model, the $\epsK$ constraint can be satisfied if
flavor changing couplings are adequately suppressed. As can be inferred from
eq. \epscon, there are roughly three different possible patterns of squark
masses that lead to small FCNCs. One is universality: scalars of the same
gauge representations get degenerate masses. This structure is generated,
at least to leading order, in many models including gauge mediation, anomaly
mediation, gaugino mediation and dilaton dominated supersymmetry breaking.
The size of deviations from universality in such theories determines how
CP violation is realized at low energies.
Another possibility for suppressing FCNCs, and therefore $\epsK$,
is through alignment of the squark mass matrices with quark mass
matrices. In such models, squark mass matrices are approximately diagonal
in the quark mass basis. Models of high scale supersymmetry breaking, in which
the squark mass matrices as well as the quark mass matrices are governed
by flavor symmetries, naturally give rise to alignment.
Finally, FCNCs, and therefore $\epsK$ are suppressed if the first
two generation sfermions are heavy.

Note that universality and alignment do not affect, in general, the
supersymmetric contributions to EDMs. On the other hand, as can be
seen
from eq. \dncon, a large mass for the first two generation sfermions will
alleviate the EDM problem.

In the next section, we will study CP violation in different models
of supersymmetry breaking which fall under these three different categories.
Since most of these models try to achieve universality, let us start by
discussing the $\epsK$ prediction in models of exact universality.

\subsec{Exact universality}
If at some high energy scale squarks are exactly degenerate and the $A$ terms
proportional to the Yukawa couplings, then the contribution to $\epsK$ comes
from RGE and is GIM suppressed, that is
\eqn\gimsup{\epsK\propto\Im(V_{td}V_{ts}^*)^2Y_t^4
\left[{\log(\Lambda_{\rm SUSY}/m_W)\over16\pi^2}\right]^2.}
This contribution is negligibly small
\ref\DGH{M. Dugan, B. Grinstein and L.J. Hall,
 Nucl. Phys. {\bf B255}, 413 (1985).}.
The contribution from genuinely supersymmetric phases ({\it i.e.} the phases
in $A_t$ and $\mu$) is also negligible
\ref\AbFr{S.~A.~Abel and J.~M.~Frere,
 Phys.\ Rev.\  {\bf D55}, 1623 (1997) [hep-ph/9608251].}%
\ref\BaKo{S. Baek and P. Ko,
 Phys. Lett. {\bf B462}, 95 (1999) [hep-ph/9904283].}.

This does not mean that there is no supersymmetric effect on $\epsK$.
In some small corner of parameter space the supersymmetric contribution
from stop-chargino diagrams can give up to 20\% of $\epsK$
\ref\BCKO{G.~C.~Branco, G.~C.~Cho, Y.~Kizukuri and N.~Oshimo,
Phys.\ Lett.\ {\bf B337}, 316 (1994) [hep-ph/9408229].}%
\ref\GNO{T. Goto, T. Nihei and Y. Okada,
 Phys. Rev. {\bf D53}, 5233 (1996) [hep-ph/9510286];
Erratum-ibid. {\bf D54}, 5904 (1996).}.

We conclude that in a supersymmetric framework with {\it nearly} exact
universality and proportionality at some high scale, the Standard Model
diagrams still have to account for $\epsK$. Note that in order to do that, the
Kobayashi-Maskawa phase has to be of order one. So in such theories,
not only is there is no motivation for approximate CP but actually this
possibility is excluded with exact universality. (It follows that the EDM
problems have to be solved by mechanisms other than approximate CP.)
We will argue below that this is the situation in models with low scale
supersymmetry breaking, but that typically degeneracy and proportionality do
not hold to such a high degree of accuracy in high scale models.

\newsec{Low Scale Supersymmetry Breaking: Gauge Mediation}
In models of Gauge Mediated Supersymmetry Breaking (GMSB),
superpartner masses are generated by the Standard Model gauge interactions.
These masses are then exactly universal at the scale $\Lambda_{\rm SUSY}$,
at which they are generated  (up to tiny high order effects associated
with Yukawa couplings). Furthermore, $A$ terms are suppressed by loop factors.
The only contribution to $\epsK$ is then from the running,
and since $\Lambda_{\rm SUSY}$ is low it is highly suppressed.

These models can also readily satisfy the EDM constraints. In most models,
the $A$ terms and gaugino masses arise from the same supersymmetry breaking
auxiliary field. They therefore carry the same phase (up to corrections
from the Standard Model Yukawa couplings), and $\phi_A$ vanishes
to a very good approximation.

The value of $\phi_B$ in general depends on the mechanism for generating
the $\mu$ term. However, running effects can generate an adequate
$B$ term at low scales in these models even if $B(\Lambda_{\rm SUSY})=0$.
One then finds
\ref\bkw{K.~S.~Babu, C.~Kolda and F.~Wilczek,
Phys.\ Rev.\ Lett.\ {\bf 77}, 3070 (1996)
[hep-ph/9605408].}
\eqn\radb{{B\over\mu} = A_t(\Lambda_{\rm SUSY})+ M_2(\Lambda_{\rm SUSY})\,
(-0.12+0.17\vert h_t\vert^2)\ ,}
where $M_2$ is the $SU(2)$ gaugino mass, $h_t$ is the top Yukawa,
and $A_t$ is the top $A$ term.
Since $\phi_A\sim0$, the resulting $\phi_B$ therefore vanishes,
again up to corrections involving the Standard Model
Yukawa couplings
\ref\dns{M.~Dine, Y.~Nir and Y.~Shirman,
Phys.\ Rev.\ {\bf D55}, 1501 (1997) [hep-ph/9607397].}.

There is therefore no CP problem in simple models of gauge mediation,
even with phases of order one. A large KM phase is actually {\it required}
in order to account for $\epsK$, so approximate CP is unacceptable.
GMSB models predict then a large CP asymmetry
in $B\rightarrow\psi K_S$, with small deviations (at most 20\%) from the SM.

This conclusion may be evaded in more complicated models
of gauge mediation. For example, matter-messenger
couplings typically spoil universality, and may carry additional
phases. The size of such couplings is of course constrained
by CP conserving flavor changing processes, but with phases
of order one they could lead to $\epsK$ values that are too large
even if the $\Delta m_K$ constraint is satisfied.
These couplings can however be forbidden by symmetries.

The fact that $\phi_A$ and $\phi_B$ vanish in simple
models of gauge mediation is really a consequence of the fact
that the soft terms in these models are generated
by a limited number of new couplings. These couplings can  often
be chosen to be real by field redefinitions, so that there are no
physical CP violating phases beyond the KM phase.
The same holds for simple models of anomaly mediated
supersymmetry breaking as we will see later.

The situation is very different in models of high scale supersymmetry breaking.
In such models, the soft terms are generated by non-renormalizable terms,
and since typically many such terms appear,
it is impossible to rotate away all phases.

\newsec{High Scale Supersymmetry Breaking}
\subsec{Dilaton dominance}
If different moduli of string theory obtain supersymmetry
breaking $F$ terms, they would typically induce flavor-dependent
soft terms through their tree-level couplings to Standard Model fields.
If however the dilaton $F$ term is the dominant one, then at tree level,
the resulting soft masses are universal and the $A$ terms proportional
to the Yukawa couplings. This is because the dilaton couplings to
Standard Model fields are constrained by the fact that the dilaton determines
the Standard Model gauge couplings.

One then finds for the gaugino masses, the scalar masses-squared,
and the $A$ term,
\eqn\treeterms{
m_{1/2}={F_S\over S+S^*}\,, \
m^2_0=\vert m_{3/2}\vert^2 = {1\over 3} {\vert F_S \vert^2\over S+S^*}\,,
\ \ A=-{F_S\over S+S^*} \,,}
where $S$ and $F_S$ are the dilaton scalar and auxiliary VEVs respectively.
There is then no new contribution to $\epsK$ in models of
dilaton dominance to leading order in string perturbation theory.

Both universality and proportionality are violated by string loop effects.
These induce corrections to squark masses
of order ${\alpha_X\over\pi}m^2_{3/2}$, where $\alpha_X=(2\pi(S+S^*))^{-1}$
is the string coupling. There is no reason why these corrections
would be flavor blind. However, RGE effects enhance the universal part of the
squark masses by roughly a factor of 5, leaving the off-diagonal entries
essentially unchanged. The flavor suppression factor is then
\ref\LoNi{J. Louis and Y. Nir,
 Nucl. Phys. {\bf B447}, 18 (1995) [hep-ph/9411429].}
\eqn\dildoe{\vev{\delta_{12}^d}\simeq
{m^{2\ {\rm one-loop}}_{12}\over m^2_{\tilde q}}
\simeq{\alpha_X\over\pi}{1\over25}\simeq4\times10^{-4}.}

Dilaton dominance relies on the assumption that loop corrections are small.
This probably presents the most serious theoretical difficulty
for this idea, because it is hard to see how non-perturbative
effects, which are probably required to stabilize the dilaton,
could do so in a region of weak coupling.  In the strong
coupling regime, these corrections could be much larger
\ref\ds{M.~Dine and M.~Graesser, {\it To appear.}}.
However, this idea at least gives some plausible theoretical explanation
for how universal masses might emerge in hidden sector models.
Given that dilaton stabilization might require that non-perturbative effects
are important, the estimate of flavor suppression \dildoe\ might well turn out
to be an underestimate.

We now turn to the flavor diagonal phases that enter
in various EDMs. From eqn.~\treeterms\ it is easy to see that
$\phi_A$ vanishes at tree-level, so that \LoNi
\ref\BIM{A.~Brignole, L.~E.~Ibanez and C.~Munoz,
Nucl.\ Phys.\  {\bf B422}, 125 (1994) [hep-ph/9308271].}
\eqn\adil{\phi_A={\cal O}\left(\alpha_X/\pi\right)\ .}
However, $\phi_B$ is unsuppressed, even when $\mu$, and through it $B$,
are generated by Kahler potential effects through supersymmetry breaking,
in which case $B=2 m_{3/2}^* \mu$
\ref\BLM{R.~Barbieri, J.~Louis and M.~Moretti,
Phys.\ Lett.\  {\bf B312}, 451 (1993) [hep-ph/9305262].}.
While the size of $m_{3/2}$ is determined from the requirement that the
cosmological constant vanishes, its phase remains arbitrary, and in fact
depends on the phase of the constant term that is added to the superpotential
in order to cancel the cosmological constant.

The most natural mechanism to suppress $\phi_B$ is then approximate CP.
Whatever physics is required to set the cosmological constant
to (essentially) zero, should obey approximate CP,
so that there is no large physical phase appearing from the gravitino mass.

We conclude that approximate CP is well motivated in models of dilaton
dominated supersymmetry breaking. Eq. \dildoe\ must, in such a case,
be an underestimate of the size of flavor changing couplings.
For EDM contributions to be small in these models, the gravitino mass
better give a small physical phase.

\subsec{Alignment}
In the framework of alignment, one does not assume any squark degeneracy.
Instead, flavor violation is suppressed because the squark mass
matrices are approximately diagonal in the quark mass basis.
This is the case in models of Abelian flavor symmetries,
in which the off-diagonal entries in both the quark
mass matrices and in the squark mass matrices are suppressed
by some power of a small parameter, $\lambda$, that quantifies
the breaking of some Abelian flavor symmetry.
A natural choice for the value of $\lambda$ is $\sin\theta_W$,
so we will take $\lambda\sim0.2$.
One would naively expect the first two generation squark mixing
to be of the order of $\lambda$.
However, the $\Delta m_K$ constraint is not satisfied with the
`naive alignment',
$K_{12}^d\sim\lambda$, and one has to construct more complicated models to
achieve the required suppression
\ref\NiSe{Y.~Nir and N.~Seiberg,
Phys.\ Lett.\  {\bf B309}, 337 (1993) [hep-ph/9304307].}%
\ref\LNS{M.~Leurer, Y.~Nir and N.~Seiberg,
Nucl.\ Phys.\  {\bf B420}, 468 (1994) [hep-ph/9310320].}.
One can also construct models where $\vev{\delta_{12}^d}\sim\lambda^5$
\ref\NiRa{Y.~Nir and R.~Rattazzi,
Phys.\ Lett.\  {\bf B382}, 363 (1996) [hep-ph/9603233].},
but these models are highly constrained and almost unique. It is simpler
to construct models where $\vev{\delta_{12}^d}\sim\lambda^3$ but the
CP violating phases are also suppressed \EyNi. In this framework,
approximate CP is then well motivated, though not unavoidable.

Note however that a basic assumption in the analyses mentioned above is
that the Abelian horizontal symmetry is the {\it only} ingredient that
plays a role in determining the squark mass-squared matrices. This assumption
does not hold if there is a large RGE contribution from gaugino masses.
If at the high scale the diagonal elements of the squark mass matrices
are comparable to the gluino mass ($m_{\tilde q}\sim{\tilde m}_{\tilde g}$),
running down to low scales enhances $m_{\tilde q}^2$ by roughly
a factor of 7. The off-diagonal elements of the squark mass
matrices remain essentially unchanged. Therefore, `naive alignment', that is,
$m^2_{12}/m_{\tilde q}^2\sim\lambda$ at high scales, would lead to
\eqn\alinai{\vev{\delta_{12}^d}\sim{\lambda\over7}\sim0.03.}

This again motivates approximate CP.
As concerns flavor diagonal phases, the question is more model dependent.
There is however a way to suppress these phases without assuming approximate
CP \NiRa. The mechanism requires that CP is spontaneously broken by the same
fields that break the flavor symmetry (``flavons"). It is based on the
observation that a Yukawa coupling and the corresponding $A$ term carry the same
horizontal charge and therefore their dependence on the flavon fields is
similar. In particular, if a single flavon dominates a certain coupling,
the CP phase is the same for the Yukawa coupling and for the corresponding $A$
term and the resulting $\phi_A$ vanishes. Similarly, if the $\mu$ term and the
$B$ term depend on one (and the same) flavon, $\phi_B$ is suppressed.

We conclude that approximate CP is well motivated in the framework of Abelian
horizontal symmetries, as it provides the most generic and easiest-to-implement
mechanism to solve both the $\epsK$ and the $d_N$ problems. One can construct
models, however, in which the horizontal symmetry solves these problems even
with phases of order one.

\subsec{Non-Abelian horizontal symmetries}
Non-Abelian horizontal symmetries can induce approximate degeneracy between
the first two squark generations, thus relaxing the flavor and CP problems
\ref\DKL{M.~Dine, R.~Leigh and A.~Kagan,
Phys.\ Rev.\  {\bf D48}, 4269 (1993) [hep-ph/9304299].}.
A review of $\epsK$ in this class of models can be found in
\ref\GNR{Y.~Grossman, Y.~Nir and R.~Rattazzi, in {\it Heavy Flavours II},
eds. A.J. Buras and M. Lindner (World Scientific Publishing Co., Singapore),
[hep-ph/9701231].}.
Quite generically, the supersymmetric contributions to $\epsK$ are too large
and require small phases (see, for example, the models of ref.
\ref\BDH{R.~Barbieri, G.~Dvali and L.~J.~Hall,
Phys.\ Lett.\  {\bf B377}, 76 (1996) [hep-ph/9512388].}).
There are however specific models where the $\epsK$ problem is solved
without the need for small phases
\ref\CHM{C.~D.~Carone, L.~J.~Hall and H.~Murayama,
Phys.\ Rev.\  {\bf D54}, 2328 (1996) [hep-ph/9602364].}%
\ref\BHRR{R.~Barbieri, L.~J.~Hall, S.~Raby and A.~Romanino,
Nucl.\ Phys.\  {\bf B493}, 3 (1997) [hep-ph/9610449].}.
Furthermore, universal contributions from RGE running might further relax
the problem. We conclude that, in the framework of non-Abelian horizontal
symmetries, approximate CP is well motivated but not unavoidable.

As concerns flavor diagonal phases, it is difficult (though not entirely
impossible) to avoid $\phi_A\gsim\lambda^2\sim0.04$ \GNR. This, however, might
be just enough to satisfy the $d_N$ constraint.

\subsec{Heavy squarks}
In models where the first two generation squarks are heavy, the basic mechanism
to suppress flavor changing processes is actually flavor diagonal:
$m_{\tilde q_{1,2}}\sim20\ {\rm TeV}$.
Naturalness does not allow higher masses, but
this mass scale is not enough to satisfy even the $\Delta m_K$ constraint
\ref\CKLN{A.~G.~Cohen, D.~B.~Kaplan, F.~Lepeintre and A.~E.~Nelson,
Phys.\ Rev.\ Lett.\  {\bf 78}, 2300 (1997) [hep-ph/9610252].},
and one has to invoke alignment, $K_{12}^d\sim\lambda$. This is still
not enough to satisfy the $\epsK$ constraint of eq. \epscon, and
approximate CP is again well motivated.

Two more comments are in order:
\item{1.} In this framework, gauginos are significantly lighter than the first
two generation squarks, and so RGE cannot induce degeneracy.

\item{2.} The large mass of the squarks is enough to solve the EDM related
problems, and so it is only the $\epsK$ constraint that motivates
approximate CP.

\subsec{Anomaly Mediation}
Another approach to solving the flavor problems of supersymmetric
theories, as well as to obtaining a predictive spectrum, is known
as Anomaly Mediation. In the presence of some truly ``hidden'' supersymmetry
breaking sector, with no couplings to the Standard Model fields (apart from
indirect couplings through the supergravity multiplet)
the conformal anomaly of the Standard Model gives
rise to soft supersymmetry breaking terms for the Standard Model fields
\ref\rs{L.~Randall and R.~Sundrum,
Nucl.\ Phys.\ {\bf B557}, 79 (1999) [hep-th/9810155].}%
\ref\Luty{G.~F.~Giudice, M.~A.~Luty, H.~Murayama and R.~Rattazzi,
JHEP {\bf 9812}, 027 (1998) [hep-ph/9810442].}.
These terms are generated purely by gravitational effects and are given by
\eqn\am{
m^2_0(\mu)=-{1\over 4}{\partial\gamma(\mu)\over\ln\mu}m^2_{3/2},\ \
m_{1/2}(\mu)= {\beta(\mu)\over g(\mu)}m_{3/2},\ \
A(\mu)=-{1\over 2}\gamma(\mu)m_{3/2},}
where $\beta$ and $\gamma$ are the appropriate beta function and anomalous
dimension. Thus, apart from the Standard Model gauge and Yukawa
couplings, the soft terms only involve the parameter $m_{3/2}$.

In general, naturalness  considerations suggest that couplings
of hidden and visible sectors should appear in the Kahler potential,
leading to soft masses for scalars already at tree level, and certainly by one
loop. As a result, one would expect the contributions \am\ to be irrelevant.

However, the authors of \rs\ argued that in ``sequestered
sector models'', in which the visible sector fields and supersymmetry
breaking fields live on different branes, separated by some distance,
the anomaly mediated contribution~\am\ could be the dominant effect.
This leads to a predictive picture for scalar masses. Since the soft
terms~\am\ are generated by the Standard Model gauge and Yukawa couplings,
they are universal, up to corrections involving the third
generation Yukawas. However, the resulting slepton masses-squared
are negative, so this model requires some modification.
We will not attempt a complete review of this subject here.
Our principal concerns are the sources of CP violation, and the
extent to which the anomaly-mediated formulae receive corrections,
leading to non-degeneracy of the squark masses.

~For eqn.~\am\ to correctly give the leading order soft terms, it is necessary
that all moduli obtain large masses before supersymmetry breaking, and that
there be no Planck scale VEVs in the supersymmetry breaking sector
\ref\bagger{J.~A.~Bagger, T.~Moroi and E.~Poppitz,
JHEP {\bf 0004}, 009 (2000) [hep-th/9911029].}.
A possible scenario for this to happen is if all moduli but the fifth
dimensional radius, $R$, sit at an enhanced symmetry point, and that $R$
obtains a large mass compared to the supersymmetry-breaking scale
(say, by a racetrack mechanism).

Even in this case, however, there is a difficulty.  One might
expect that some of the moduli have masses well below the fundamental scale.
If there are light moduli in the bulk, there are typically one-loop
contributions to scalar masses squared from exchanges of
bulk fields, proportional to $m_{3/2}^2/R^3$ times a loop factor~\rs.\foot{
There are potentially even larger corrections.  At tree level,
in the Horava-Witten picture, there are tadpoles for bulk fields,
and these can easily lead to contributions which swamp those of
\am.  These issues are under investigation and will be reported elsewhere.}
Indeed, the authors of ref. \rs\ suggested that these contributions were
universal, and could provide the necessary non-tachyonic contributions to
scalar masses.  We do not understand, however, the statement that
these contributions are  generically universal.  In the case of, say,
Horava-Witten theory compactified on a Calabi-Yau space, the
couplings of the matter fields to the different bulk moduli are
in general not the same. Since minimal AMSB gives
$\tilde m_{\rm slepton}^2 \sim 0.05\, \tilde m_{\rm squark}^2$,
such non-universal contributions, if responsible for the positive part of the
slepton masses, would violate the $\Delta m_K$ constraint (see~\ranmot).

It is possible to suppress these contributions by taking $R$ to be large.
To satisfy the $\Delta m_K$ constraint, one would need $R\gsim 30$,
for which the 4d Planck scale is about a factor of 5 above the 5d Planck scale.
In this case, approximate CP is well motivated. Large values of $R$, however,
might be problematic, given Witten's observation
\ref\witten{E.~Witten,
Nucl.\ Phys.\ {\bf B471}, 135 (1996) [hep-th/9602070].}
that generically for large $R$ the gauge couplings are proportional to $1/R$.

If there are no light moduli, and if the contributions described above are
adequately suppressed, some modification of the visible sector is required
in order to generate acceptable slepton masses. Different such solutions
have been suggested
\nref\pora{A.~Pomarol and R.~Rattazzi,
JHEP {\bf 9905}, 013 (1999) [hep-ph/9903448].}%
\nref\kss{E.~Katz, Y.~Shadmi and Y.~Shirman,
JHEP {\bf 9908}, 015 (1999) [hep-ph/9906296].}%
\nref\clpss{Z.~Chacko, M.~A.~Luty, E.~Ponton, Y.~Shadmi and Y.~Shirman,
hep-ph/0006047.}%
\nref\clmp{Z.~Chacko, M.~A.~Luty, I.~Maksymyk and E.~Ponton,
JHEP {\bf 0004}, 001 (2000) [hep-ph/9905390].}%
\nref\jajo{I.~Jack and D.~R.~Jones, Phys.\ Lett.\ {\bf B491}, 151 (2000)
[hep-ph/0006116];
Phys.\ Lett.\ {\bf B482}, 167 (2000) [hep-ph/0003081].}%
\nref\alde{B.~C.~Allanach and A.~Dedes,
JHEP {\bf 0006}, 017 (2000) [hep-ph/0003222].}%
\refs{\pora-\alde}. In some of these models,
there are no new contributions to CP violation
simply because there are few enough new parameters in the theory
that they can all be chosen real by field redefinitions~\pora\kss\clpss.
Furthermore, it is possible to generate the $\mu$ term in these
models from AMSB, so that $\phi_B$ vanishes. These models are then similar to
GMSB models from the point of view of CP violation.
Approximate CP is not only unmotivated in these models,
it is actually ruled out by $\epsK$. Whether such a situation can actually be
realized in a fundamental theory is unclear.

We conclude then, that in generic sequestered sector models
it is difficult to obtain large degeneracy; approximate CP is well motivated.
It is conceivable that there might be theories with a high degree of
degeneracy, or with no new sources of CP violation.
In such theories, approximate CP would be ruled out.

\subsec{Gaugino Mediation}
Gaugino mediation
\ref\schmaltz{D.~E.~Kaplan, G.~D.~Kribs and M.~Schmaltz,
Phys.\ Rev.\ D {\bf 62}, 035010 (2000) [hep-ph/9911293].}%
\ref\ann{Z.~Chacko, M.~A.~Luty, A.~E.~Nelson and E.~Ponton,
JHEP {\bf 0001}, 003 (2000) [hep-ph/9911323].}\
is in many ways similar to anomaly mediation, and poses similar issues.
These models also suppress dangerous tree level contact terms by
invoking extra dimensions, with the Standard Model matter fields localized
on one brane and the supersymmetry breaking sector on another brane.
In this case, however, the Standard Model gauge fields are in the
bulk, so gauginos get masses at tree level, and as a result
scalar masses are generated by running. Scalar masses are therefore universal.
Furthermore, the soft terms typically involve only one new
parameter, namely, the singlet $F$ VEV that gives rise
to gaugino masses. Therefore, they do not induce any new CP violation.
If there are no additional contributions to the soft terms,
there is no motivation for approximate CP.

Again, however, if there are non-universal tree and one loop contributions
to scalar masses, as in the Horava-Witten picture, significant violations of
degeneracy and proportionality can be expected, and approximate CP
is well motivated.

\newsec{The $\epe$ Problem}
If all phases and, in particular, the KM phase, are small, then the Standard
Model cannot account for $\epe$. Consequently, not only $\epsK$ but also $\epe$
have to be accounted for by supersymmetric contributions.

A typical supersymmetric contribution to $\epe$ is given by
\ref\BCIRS{A.~J.~Buras, G.~Colangelo, G.~Isidori, A.~Romanino and
L.~Silvestrini, Nucl.\ Phys.\  {\bf B566}, 3 (2000) [hep-ph/9908371].}
\eqn\susyepe{\eqalign{|\epe|=58B_G\ &\ \left[{\alpha_s(m_{\tilde g}\over
\alpha_s(500\ {\rm GeV})}
\right]^{23/21}\left({158\ MeV\over m_s+m_d}\right)\cr
\times&\ \left({500\ {\rm GeV}\over
m_{\tilde g}}\right)\left|\Im\left[(\delta_{LR}^d)_{12}-(\delta_{LR}^d)_{21}^*
\right]\right|.\cr}}
Consequently, the supersymmetric contribution saturates $\epe$ for
\eqn\satepe{\Im\left[(\delta_{LR}^d)_{12}-(\delta_{LR}^d)_{21}^*\right]\
\sim\ \lambda^7\left({m_{\tilde g}\over500\ {\rm GeV}}\right)}
where, motivated by flavor symmetries, we parameterize the suppression
by powers of $\lambda\sim0.2$.
(Note that the related hadronic uncertainties are large, and a very
conservative estimate would give $\lambda^9$ as a lower bound.)

Without proportionality, a naive guess would give
\eqn\naiepe{\eqalign{
(\delta_{LR}^d)_{12}\ \sim&\ {m_s|V_{us}|\over\tilde m}\sim\lambda^{5-6}
{m_t\over\tilde m},\cr
(\delta_{LR}^d)_{21}\ \sim&\ {m_d\over|V_{us}|\tilde m}\ \sim\lambda^{5-6}
{m_t\over\tilde m}.\cr}}
This is not far from the value required to account for $\epe$
\ref\MaMu{A.~Masiero and H.~Murayama,
Phys.\ Rev.\ Lett.\  {\bf 83}, 907 (1999) [hep-ph/9903363].}.
With small phases of order $\lambda^2$, the situation is still very promising.
The problem is, however, that eq. \naiepe\ gives an overestimate of the
supersymmetric contribution in viable models of supersymmetry breaking
that have appeared in the literature.

With dilaton dominance, the $A$ terms are proportional to the Yukawa terms
at tree level, so that there is a suppression of ${\cal O}(\alpha_X/\pi)\sim
10^{-2}$ compared to \naiepe.

With alignment, the $\epsK$ constraint requires that the relevant terms
are suppressed by at least a factor of $\lambda^2$ compared to \naiepe\
\ref\EMNS{G.~Eyal, A.~Masiero, Y.~Nir and L.~Silvestrini,
JHEP {\bf 9911}, 032 (1999) [hep-ph/9908382].}.

With heavy squarks, a more likely estimate is \EMNS
\eqn\hqepe{(\delta_{LR}^d)_{ij}\ \sim\ {m_Z(M_d)_{ij}\over(10\ {\rm TeV})^2},}
which suppresses the relevant matrix elements by a factor of order $10^4$
compared to \naiepe. (One can perhaps construct models with enhanced $A$ terms,
resulting in $(\delta_{LR}^d)_{ij}\sim(M_d)_{ij}/(10\ {\rm TeV})$, but even in
this case there is a suppression of order $10^{-2}$ compared to \naiepe.)

With a horizontal U(2) symmetry, the two contributions in \susyepe\ cancel
each other. (More generally, this happens for a symmetric $A$ matrix
with $A_{11}=0$
\ref\BCS{R.~Barbieri, R.~Contino and A.~Strumia,
Nucl.\ Phys.\  {\bf B578}, 153 (2000) [hep-ph/9908255].}.)

It seems then that generically, in all the frameworks discussed above,
approximate CP is inconsistent with $\epe={\cal O}(10^{-3})$ (see also
\AbFr). There are three possibilities:
\item{1.} The hadronic parameters (particularly $B_G$ and $m_s$) are at
the extreme of their `reasonable ranges' and supersymmetry with flavor
suppression of order $10^{-1}-10^{-2}$ compared to \naiepe\ does account
for $\epe$ \EMNS.
\item{2.} Approximate CP is not realized in nature. The CP problems
of high energy supersymmetry breaking may be solved in a different way
(see {\it e.g.}
\ref\MaVi{A. Masiero, M. Piai and O. Vives, hep-ph/0012096.}).
Alternatively the scale of supersymmetry breaking could be low.
\item{3.} The structure of the LR block in the squark mass-squared matrices
is different from all models discussed above.

As an example of the third possibility, consider the proposal of ref.
\ref\BJKP{S.~Baek, J.~H.~Jang, P.~Ko and J.~H.~Park,
Phys.\ Rev.\  {\bf D62}, 117701 (2000) [hep-ph/9907572].}.
It is shown there that it is possible to account for both $\epsK$ and $\epe$
by supersymmetric contributions with (i) $|(\delta^d_{LL})_{12}|\sim
10^{-3}-10^{-2}$ with a phase of ${\cal O}(1)$; and
(ii) $|(\delta^d_{LR})_{22}|\sim10^{-2}$.
The first factor accounts for $\epsK$ and the product of the two insertions
for $\epe$. Note that the value of $|(\delta^d_{LR})_{22}|$ is much larger than
our naive estimate, $m_s/\tilde m\sim10^{-3}$. The proposed mechanism is to
have a very large $\tan\beta$, in which case it is the contribution proportional
to $\mu\tan\beta$ (rather than to $A^d$) that dominates $\delta^d_{LR}$.

Can this mechanism be modified to the case of approximate CP? For
$\delta^d_{LR}$, approximate CP is actually useful since, to satisfy the $d_N$
constraint, the large value of $\mu\tan\beta$ requires a very small $\phi_B$.
As concerns the $(\delta^d_{LL})$, ref. \BJKP\
considers a phase of order one. But the phase can be taken to be small with
the absolute value as large as allowed by the $\Delta m_K$ constraint.
The main problem is that a very strong suppression of $|(\delta^d_{RR})_{12}|$
is required.

Another possibility is that the flavor symmetry is an R symmetry, which
would generate a different structure in $M^d$ and $A^d$.

We emphasize that there is no fundamental reason that the $A$-terms
would not saturate \naiepe. The fact that this does not happen in existing
models of approximate CP only means that finding an attractive model of
approximate CP that is consistent with the $\epe$ constraints is still
an important task for model builders.

\newsec{Conclusions}
All measured CP violations, namely the $\epsK$ and $\epe$ parameters
in neutral $K$ decays, are small. Recent measurements of the CP
asymmetry in $B\rightarrow\psi K_S$
\ref\cdf{T. Affolder {\it et al.}, CDF collaboration,
Phys. Rev. {\bf D61}, 072005 (2000)  [hep-ex/9909003].}%
\ref\babar{D. Hitlin, BaBar collaboration, plenary talk
in ICHEP 2000 (Osaka, Japan, July 31, 2000), SLAC-PUB-8540.}%
\ref\belle{H. Aihara, Belle collaboration, plenary talk
in ICHEP 2000 (Osaka, Japan, July 31, 2000).}
yield an average of $a_{\psi K_S}=0.42\pm0.24$, which
still leaves open the possibility that this asymmetry is also small
\ref\ENP{G.~Eyal, Y.~Nir and G.~Perez,
JHEP {\bf 0008}, 028 (2000) [hep-ph/0008009].}.

The Standard Model explains the smallness of $\epsK$ and $\varepsilon^\prime$
as a result of small flavor changing couplings, while the Kobayashi-Maskawa
phase is {\it required} to be large. The large phase leads to a prediction
that $a_{\psi K_S}$ is large, roughly $a_{\psi K_S}\sim0.5-0.9$. The same
situation generically applies in supersymmetric models with a low
breaking scale, such as gauge mediated supersymmetry breaking.

Supersymmetric models with a high breaking scale often give unacceptably
large contributions to EDMs and have insufficient flavor suppression for
$\epsK$. In such models,
approximate CP, that is the smallness of all CP violating phases, is
therefore well motivated. This applies to models of dilaton
dominance, Abelian and non-Abelian flavor symmetries, heavy squarks
and very likely also to anomaly-mediated and gaugino-mediated
supersymmetry breaking. The smallness of $\epsK$ and $\epe$ is then explained
by the smallness of the phases. In contrast to the Standard Model and to
GMSB, CP asymmetries in $B$ decays and, in particular, $a_{\psi K_S}$
are predicted to be small.

In the near future, $a_{\psi K_S}$ will be experimentally determined
with much better accuracy. It could be that the value of the asymmetry
will be found to lie within the Standard Model range. In such a case,
the idea of approximate CP will be excluded. Within the supersymmetric
framework, low energy breaking will be favored. Models with high breaking
scale will not be excluded, but the CP problems of these models will become
even more puzzling.

It could also be that the value of $a_{\psi K_S}$ will be found to be small.
In such a case, the Standard Model will be excluded together with simple
GMSB models. Models of high scale supersymmetry breaking will be favored.
Finding attractive mechanisms for producing large enough $\epe$ and improving
existing suggestions for inducing approximate CP from fundamental theories will
become important challenges.

\vskip 0.5cm

\centerline{\bf Acknowledgments}
We thank Alon Faraggi, Michael Graesser and Scott Thomas for 
useful discussions.
This research project is supported by the United States $-$
Israel Binational Science Foundation (BSF). 
The work of M.D. is supported in part by the U.S. Department of Energy.
Y.N. is supported by the Israel Science Foundation founded by the
Israel Academy of Sciences and Humanities and by the Minerva
Foundation (Munich).  

\listrefs
\end